\documentclass[twocolumn,aps, prd]{revtex4-2}
\usepackage[colorlinks,linkcolor=blue,citecolor=blue,bookmarks,pdfstartview=FitH]{hyperref}
\usepackage{graphicx,epsf}
\usepackage{color}
\usepackage{dcolumn}
\usepackage{amsfonts}
\usepackage{graphics}
\usepackage{epstopdf}
\usepackage{mathrsfs}
\usepackage[mathscr]{eucal}
\usepackage{amsmath,amsfonts,amssymb,bm}
\usepackage{upgreek}
\usepackage{caption}
\makeatletter
\def\tagform@#1{\maketag@@@{\ignorespaces#1\unskip\@@italiccorr}}
\let\orgtheequation\theequation
\def\theequation{(\orgtheequation)}

\usepackage{subfig}

\begin{document}

\title{On How Zonal Fields Suppress Reversed Shear Alfv\'en Eigenmode in Tokamak Plasmas}

\author{
  Ruirui Ma\textsuperscript{1,2},
  Pengfei Liu\textsuperscript{3},
  Liu Chen\textsuperscript{4,2,*},
  Fulvio Zonca\textsuperscript{2,4},
  Zhiyong Qiu\textsuperscript{5,2}
}
\affiliation{
  \footnotesize
  \noindent
  \textsuperscript{1}Southwestern Institute of Physics, P.O. Box 432 Chengdu 610041, People's Republic of China\\
  \textsuperscript{2}Center for Nonlinear Plasma Science and C.R. ENEA Frascati, Via E. Fermi 45, 00044 Frascati, Italy\\
  \textsuperscript{3}Institute of Physics, Chinese Academy of Sciences, Beijing 100190, People's Republic of China\\
  \textsuperscript{4}Institute for Fusion Theory and Simulation and School of Physics, Zhejiang University, Hangzhou 310027, People's Republic of China\\
  \textsuperscript{5}Institute of Plasma Physics, Chinese Academy of Sciences, Hefei 230031, People's Republic of China}

\date{\today}
\begin{abstract}
Employing both nonlinear gyrokinetic simulations and theoretical analyses, we have discovered the novel result that, with energetic particle dynamics kept linear, the nonlinear suppression and eventual saturation of reversed-shear Alfv\'en eigenmode occur via the downward frequency chirping induced by the beat-driven zonal current. More specifically, as the mode frequency chirps downward, there is enhanced mode conversion to radially propagating electron Landau-damped kinetic Alfv\'en waves; resulting in enhanced convective (radiative) damping and, thereby, its suppression and saturation. Theoretical results are in good agreement with simulations both qualitatively and quantitatively.
\end{abstract}


\maketitle

\newpage

{\it Introduction.}$-$In tokamak reactors, achieving high-performance plasma confinement is essential for efficient fusion energy production. Reversed-shear Alfv\'en eigenmodes (RSAEs) \cite{Kimura1998,Sharapov2001}, ubiquitous in advanced scenarios with non-monotonic safety factor profiles \cite{Liu2022}, however, can destabilize the plasma and lead to significant losses of energetic particles (EPs), which are crucial for sustaining fusion reactions. A comprehensive understanding of RSAE dynamics, particularly the mechanisms governing their nonlinear evolution and eventual saturation, is therefore crucial for optimizing plasma performance. A key candidate for mediating this saturation is zonal electromagnetic fields (ZFs), including zonal flows and zonal currents. ZFs are prevalent self-organized phenomena from planetary atmospheres to laboratory plasmas. In tokamaks, these toroidally symmetric, zero-frequency corrugations of plasma equilibria play a fundamental role in regulating microturbulence and mediating cross-scale couplings \cite{Chen2001,Diamond2005,Chen2012}. Understanding how ZFs govern RSAE nonlinear dynamics has been a crucial challenge in fusion plasma physics, and motivated the present work, where we discover a novel nonlinear saturation mechanism based on zonal-current-induced downward frequency chirping. Our gyrokinetic simulations and theoretical analyses show that this chirping enhances mode conversion to electron Landau-damped kinetic Alfv\'en waves (KAWs) \cite{Hasegawa1975,Hasegawa1976}, resulting in dramatically increased convective (radiative) damping. This paradigm shift reveals zonal-current-mediated energy channeling as one of the fundamental saturation mechanisms for Alfv\'en eigenmodes in burning plasmas.

Numerical simulations have established that self-interactions can lead to saturation of EP-driven RSAEs \cite{Cheny2018,Wang2024,Liu2023} via two main routes \cite{Chen2013a}: one involving nonlinear thermal plasma dynamics \cite{Wei2021,Wang2024} yielding nonlinear beat-driven ZFs, and the other producing long-lived nonlinear modification of the EP phase-space structures, which are manifestation of EP nonlinear equilibria in the presence of a finite level of fluctuations and are called phase-space zonal structures (PSZS) \cite{Zonca2015a,Zonca2021a,Falessi2023}. The first route is typically responsible of nonlinear frequency shift and/or modification of the local current/safety factor profile, thereby enhancing continuum damping \cite{Wei2021,Wang2024}; while the second route is responsible for the nonlinear modification of the EP driving rate \cite{Zonca2015a,Zonca2021a,Falessi2023}. Contrary to conventional expectations, recent simulation experiments and theoretical analyses \cite{Chen2025} reveal that ZFs can enhance the EP drive of RSAEs, rather than suppressing it, suggesting that ZFs-induced suppression primarily takes place through the thermal plasma route. However, present understanding of this process relies solely on large-scale numerical simulations, lacking a first-principles theoretical framework capable of predicting the dominant nonlinear effect and its underlying physical mechanisms.

This Letter investigates the roles of ZFs due to the nonlinear dynamics of thermal plasmas only in the suppression and eventual saturation of a single-$n$ RSAE. Nonlinear gyrokinetic simulations of the RSAE interacting with beat-driven ZFs exhibit clear downward frequency chirping, excitation of KAWs, and ultimate saturation. To understand the underlying physics, we formulate the RSAE eigenmode equations by self-consistently including both zonal flow and zonal current in the magnetohydrodynamic (MHD) framework. Analytical investigation reveals zonal current as the dominant mechanism responsible for downward frequency chirping, while zonal flow yields only a minor upward frequency shift. Theory further shows that Alfv\'en continuum resonant-damping is negligible. As the frequency approaches the shear Alfv\'en wave continuum, enhanced coupling to radially-propagating KAWs introduces significant convective (radiative) damping. To illustrate these processes, we develop an integrated `RSAE-ZF-KAW' coupling model that incorporates both finite Larmor radius (FLR) and electron-Landau damping dynamics, which are both indispensable for the proper description of KAW propagation and absorption. In order to quantitatively demonstrate this crucial point, we also solve a corresponding `RSAE-ZF-MHD' model, where the shear Alfv\'en continuous spectrum is considered in the ideal MHD limit. Numerical solutions of the eigenvalue equations for the `RSAE-ZF-MHD' and `RSAE-ZF-KAW' models verify the analytically predicted downward chirping induced by zonal current and quantitatively demonstrate the essential role of radiative damping in the saturation of RSAE. Our theoretical model, furthermore, demonstrates the importance of accounting for electron-Landau damping in the description of KAW dynamics; illuminating, thus, a gap in present gyrokinetic simulations neglecting it. Even further, it also provides a general paradigm for analyzing Alfv\'en-eigenmodes-ZF interactions in toroidal plasmas.

{\it Nonlinear Gyrokinetic Simulations of RSAE}.$-$We employ the Gyrokinetic Toroidal Code (GTC) \cite{Lin1998} to investigate how ZFs suppress RSAE through thermal plasma nonlinearity. Adopting DIII-D discharge \#159243 parameters at 805 ms \cite{Collins2016} featuring reversed magnetic shear (with minimum safety factor $q_{\text{m}}=2.947$), we simulate an $n=4$ RSAE, retaining full nonlinearity for thermal plasmas to isolate this route while keeping EP dynamics linear. The simulation setup follows the well-established configuration of Refs. \cite{Liu2024,Chen2025}, utilizing a global field-aligned mesh that resolves kinetic scales, a low-noise $\delta f$ scheme, and a time step capable of capturing both wave dynamics and electron motion. Electrons are treated with a fluid-kinetic hybrid electron model \cite{Lin2001}, while ions are fully gyrokinetic.

\begin{figure}[ht]
\centering
\includegraphics[width=0.31\textwidth]{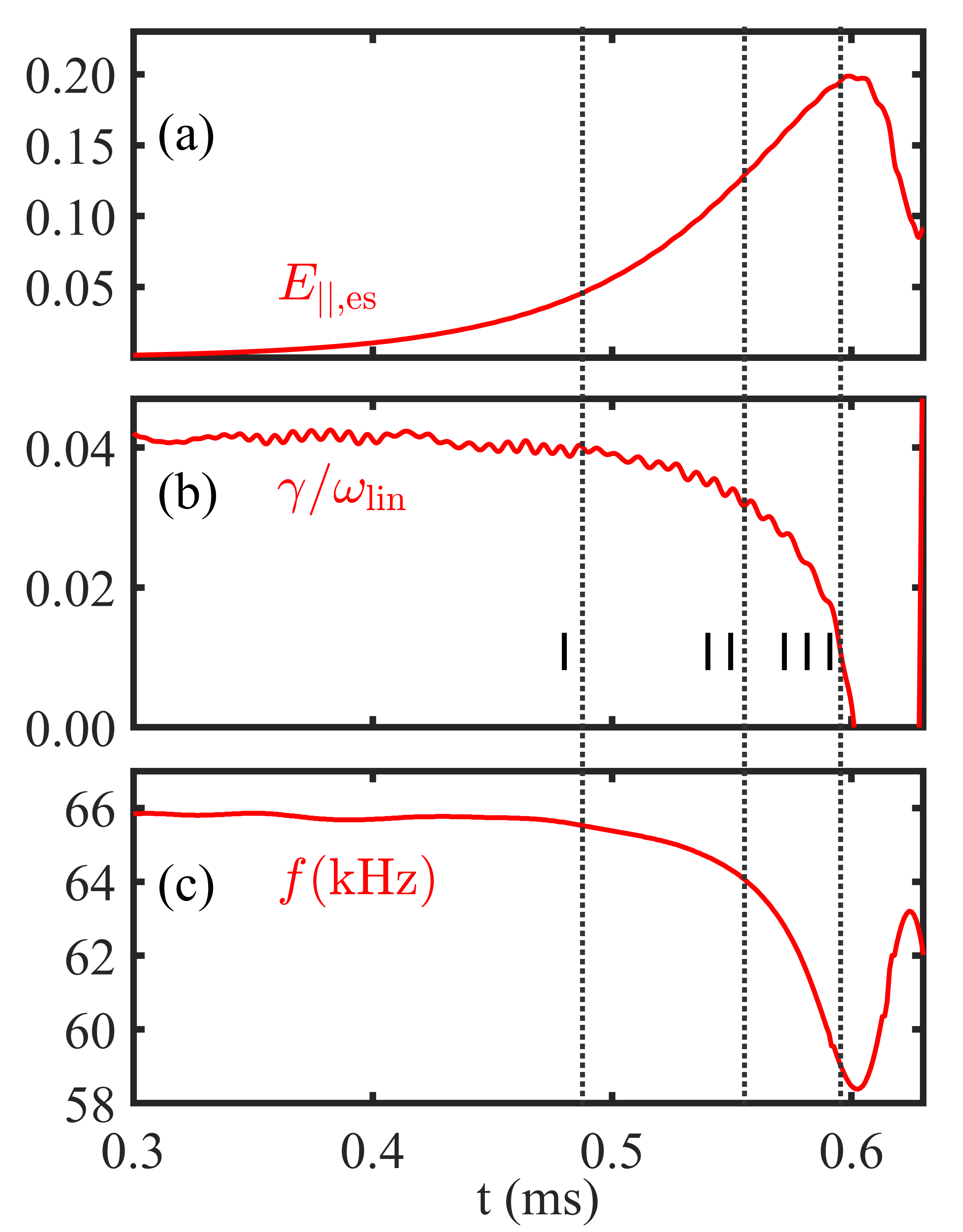}
\caption{Saturation of the $n=4$ RSAE in GTC simulation with nonlinear thermal plasma and linear EP response. Time evolution of (a) normalized amplitude of the electrostatic parallel electric field $E_{\parallel,{\rm es}}=-\nabla_\parallel\delta\phi$, (b) normalized growth rate $\gamma/\omega_{\rm lin}$ for the dominant $m=12$ poloidal harmonic, and (c) instantaneous frequency $f$ (kHz) at the $q_{\rm m}$ surface. Vertical dot lines (I, II, III) mark distinct phases of the evolution, and $\omega_{\rm lin}=2\pi\times65.8$ kHz is the RSAE linear frequency.}
\label{gtc_sim_fig1}
\end{figure}

\begin{figure}[ht]
\centering
\includegraphics[width=0.42\textwidth]{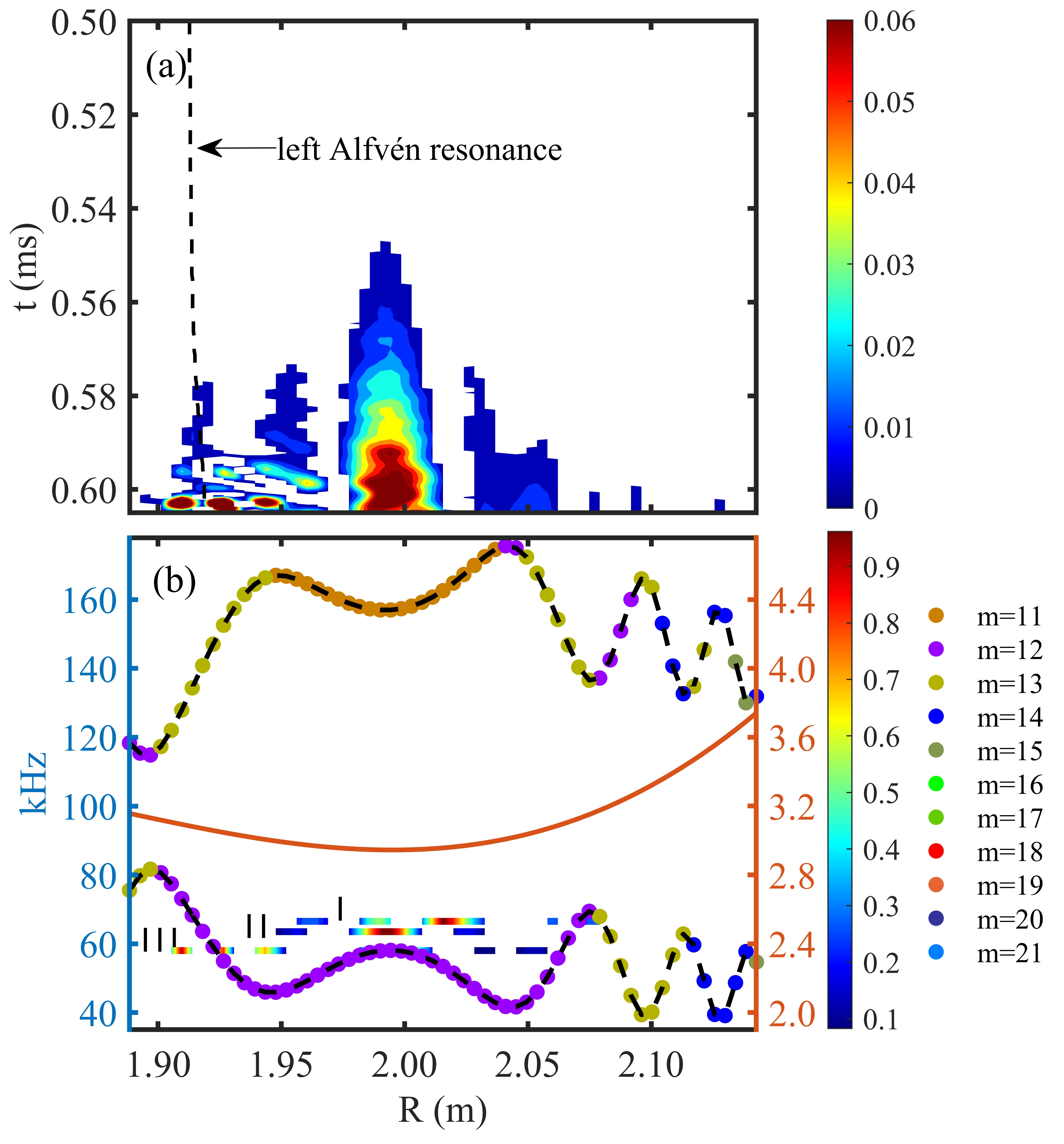}
\caption{KAW excitation and RSAE frequency evolution. (a) Time evolution of $E_\parallel$ radial profile, with Alfv\'en resonance inside $q_{\rm m}$ (dashed). (b) RSAE frequency relative to Alfv\'en continua from MAS \cite{Bao2023}; $q$ profile (orange, right axis). Horizontal lines show $E_\parallel$ structure at times I, II, III from GTC simulation (FIG. \ref{gtc_sim_fig1}).}
\label{gtc_sim_fig2}
\end{figure}

\begin{figure}[htbp]
\centering
\includegraphics[width=0.35\textwidth]{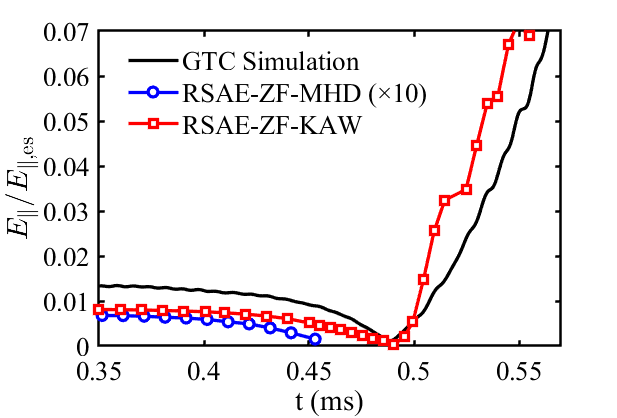}
\caption{Quantitative comparison of the evolution of $E_\parallel/E_{\parallel,{\rm es}}$: GTC Simulation (black), `RSAE-ZF-MHD' model (Eq. \ref{AE_ZF_eigen_eq}, blue circles, scaled $\times 10$), and `RSAE-ZF-KAW' model (Eq. \ref{RSAE_KAW_e_Landau_model}, red squares).}
\label{RSAE_KAW_deltaE_parallel}
\end{figure}

Our simulations demonstrate, in the present study, RSAE saturation mediated solely by thermal plasma nonlinearities. Even with linear EP drive, the mode saturates at finite amplitude (FIG. \ref{gtc_sim_fig1}(a)), exhibiting three key signatures of the underlying mechanism: suppression of the nonlinear growth rate (FIG. \ref{gtc_sim_fig1}(b)), a pronounced downward frequency chirping (FIG. \ref{gtc_sim_fig1}(c)), and the excitation of KAWs. The KAW excitation is evidenced by a rapid, order-of-magnitude increase in the parallel electric field ratio $E_\parallel/E_{\parallel,{\rm es}}$ (with $E_{\parallel} = -(\nabla_{\parallel}\delta\phi + \partial_t \delta A_{\parallel}/c)$), as is quantitatively compared with theoretical predictions in FIG. \ref{RSAE_KAW_deltaE_parallel}, and by the resulting KAWs propagating radially inward toward the magnetic axis (FIG. \ref{gtc_sim_fig2}(a)). This mode conversion is triggered as the downward-chirping RSAE frequency approaches the accumulation point at $q_{\rm m}$ of the shear Alfv\'en wave continuum (FIG. \ref{gtc_sim_fig2}(b)), opening an efficient channel for energy dissipation.

Below, in order to elucidate the underlying physics, we construct first-principles theoretical models to quantify the dominant damping mechanism and distinguish the respective roles of zonal current and zonal flow.

{\it RSAE-ZF-MHD Model}.$-$We first analyze RSAE dynamics in the presence of beat-driven ZFs within the MHD framework, considering a large aspect-ratio tokamak ($\epsilon\equiv r/R\sim {\cal O}(10^{-1}) \ll 1$) with circular magnetic surfaces. Following the theoretical approach of Ref. \cite{Chen2012}, the fluctuating fields are decomposed into RSAE components ($\delta\phi_0$,  $\delta A_{\parallel 0}$) with toroidal mode number $n_0$ and zonal components ($\delta\phi_z$,  $\delta A_{\parallel z}$):
\begin{equation}\label{delta_phi_0}
\delta\phi_0=e^{-i\omega_{0r}t+in_0\xi}\sum_m\Phi_m e^{-im\theta}+\text{c.c.},
\quad \delta\phi_z=\Phi_z(r, t), \nonumber
\end{equation}
with analogous expressions for $\delta A_\parallel$. Here, $\xi$, $\theta$, and $r$ are the toroidal-angle, poloidal-angle, and radial coordinates, respectively.

The governing equations are the nonlinear gyrokinetic equation \cite{Frieman1982}, quasi-neutrality condition, and gyrokinetic vorticity equation \cite{Chen2016}. The particle non-adiabatic perturbed distribution function for each particle species $j$ is decomposed into linear ($\delta g_j^{(1)}$) and nonlinear ($\delta g_j^{(2)}$) components. For electrons, noting $|\omega_0|,|\omega_{de}|_0\ll|k_\parallel\upsilon_{te}|_0$ and $|k_\perp\rho_e|^2\ll 1$, we obtain the massless electron responses;
\begin{equation}\label{delta_g_e0_1_2}
\begin{aligned}
&\delta g_{e0}^{(1)}\simeq -\frac{e}{T_e}F_{Me}\left(1-\frac{\omega_{\ast e}}{\omega}\right)_0\delta\psi_0,\\
&\delta g_{e0}^{(2)}\simeq -i\frac{c}{B_0}\frac{k_{\theta 0}k_z}{\omega_{0r}}\left(\delta\psi_0\delta g_z-\delta\psi_z\delta g_0\right)_e,
\end{aligned}
\end{equation}
where $\delta\psi_0=\omega_0\delta A_{\parallel 0}/ck_{\parallel 0}$ and $\delta\psi_z=\omega_0\delta A_{\parallel z}/ck_{\parallel 0}$. Meanwhile, for thermal ions with $|\omega_0| \gg |k_\parallel\upsilon_{ti}|_0$, we have
\begin{equation}\label{delta_g_i0_1_2}
\begin{aligned}
&\delta g_{i0}^{(1)}\simeq \frac{e}{T_i}F_{Mi}\left(1-\frac{\omega_{\ast i}}{\omega}\right)_0 J_0 \delta\phi_0\left(1+\frac{\omega_{di}}{\omega}\right)_0, \\
&\delta g_{i0}^{(2)}\simeq i\frac{c}{B_0}\frac{k_{\theta 0}k_z}{\omega_{0r}}\left(J_z\delta\phi_z\delta g_0-J_0\delta\phi_0\delta g_z\right)_i.
\end{aligned}
\end{equation}
Here, $\delta g_{ez}$ and $\delta g_{iz}$, respectively, denote the electron and ion responses to the ZFs, with their detailed derivations provided in Refs. \cite{Chen2025}. Substituting the $\delta g_j$'s expressions given by Eqs. \ref{delta_g_e0_1_2}-\ref{delta_g_i0_1_2} into the quasi-neutrality condition yields, after some straightforward algebra, the nonlinear Ohm's law \cite{Chen2012}:
\begin{equation}\label{nl_Ohms_law}
\delta\phi_0-\delta\psi_0\simeq i\frac{c}{B_0}\frac{k_{\theta 0}k_z}{\omega_{0}}\delta\phi_0(\delta\phi_z-\frac{\omega_0\delta A_{\parallel z}}{ck_{\parallel 0}}).
\end{equation}
Equation \ref{nl_Ohms_law} describes the ZF-induced finite parallel electric field. In this MHD description, the FLR effects ($\propto k_\perp^2\rho_i^2$) in Eq. \ref{nl_Ohms_law} are neglected; these will be incorporate when we consider the mode conversion to KAWs \cite{Hasegawa1975,Hasegawa1976}.

The other field equation, the gyrokinetic vorticity equation \cite{Chen2016},
which combines the ${\bm\nabla}\cdot\delta{\bm J}=0$ and parallel Amp\`ere's law $\nabla_\perp^2\delta A_\parallel=-4\pi\delta J_\parallel/c$, can be cast as
\begin{equation}\label{diverge_j}
{\bm\nabla}\cdot \delta{\bm J}_k^{(1)}+{\bm\nabla}\cdot\delta{\bm J}_k^{(2)}=0,
\end{equation}
where ${\bm\nabla}\cdot\delta{\bm J}_0^{(1)}$ is the well-known linear response for the RSAE:
\begin{equation}\label{diverge_j0_1}
\begin{aligned}
{\bm\nabla}&\cdot\delta{\bm J}_0^{(1)}=\frac{\delta{\bm B}_{\perp 0}}{B_0}\cdot{\bm\nabla}J_{\parallel 0}-ik_{\parallel 0}\left(\frac{c}{4\pi}\right)\nabla_\perp^2\delta A_{\parallel 0} \\
&+i\frac{c^2}{4\pi}{\bm\nabla}_\perp\cdot\frac{(\omega_0-\omega_{\ast pi})}{\upsilon_{\rm A}^2}{\bm\nabla}_\perp\delta\phi_0 \\
&+i\frac{c^2}{4\pi}{\bm\nabla}_\perp\cdot\frac{(1-\omega_{\ast pi}/\omega_0)}{\upsilon_{\rm A}^2}\frac{\omega_{\rm gam}^2}{\omega_0}{\bm\nabla}_\perp\delta\phi_0+\lambda_a,
\end{aligned}
\end{equation}
where the terms represent, respectively, the kink drive, field-line-bending, ion inertia including diamagnetic drift, and the last two curvature-pressure couplings terms, with $\lambda_a=-\frac{c}{B_0}{\bm b_0}\times{\bm\kappa}\cdot{\bm\nabla}\big[\sum_j\big\langle\big(\frac{em}{T}F_{M}\big)_j\frac{\omega_{\ast pj}}{\omega_0}\big(\frac{\upsilon_\perp^2}{2}+\upsilon_\parallel^2\big)_j\big\rangle_\upsilon\big]_s\delta\phi_0$ representing the favorable average curvature. Here, $\langle...\rangle_\upsilon$ and $\left[...\right]_s$ denote velocity-space integration and the flux-surface averaging, respectively \cite{Zonca2002}, ${\bm b}_0={\bm B}_0/B_0$, ${\bm\kappa}={\bm b}_0\cdot{\bm\nabla b}_0$, $\omega_{\ast pj}=(cT/eB_0)_j({\bm k}\times{\bm b}_0)\cdot{\bm\nabla}\ln P_j$ for species $j$, and $\omega_{\rm gam}^2=(\upsilon_{ti}/R)^2(\tau+7/4)$ is the geodesic-acoustic frequency \cite{Winsor1968,Turnbull1993,Zonca1996}. For clarity, Eq. \ref{diverge_j0_1} retains only the reactive components, neglecting wave-particle resonances in the present analysis.

${\bm\nabla}\cdot\delta{\bm J}_k^{(2)}$ in Eq. \ref{diverge_j} corresponds to nonlinear contributions due to the Maxwell stress and gyrokinetic Reynold stress \cite{Chen2016}; that is, for RSAE,
\begin{equation}\label{diverge_j_2}
\begin{aligned}
{\bm\nabla}\cdot\delta{\bm J}_0^{(2)}\!=\!\frac{c^3k_{\theta 0}k_z}{4\pi B_0\upsilon_{\rm A}^2}\big(k_{z}^2&-k_{\perp 0}^2\big)\!\bigg[\left(\frac{k_{\parallel 0}\upsilon_{\rm A}}{\omega_0}\right)\!\left(\frac{\upsilon_{\rm A}}{c}\right)\delta A_{\parallel z}\\
&-\left(1-\frac{\omega_{\ast pi}}{\omega_0}\right)\delta\phi_z\bigg]\delta\phi_0.
\end{aligned}
\end{equation}

The combination of Eqs. \ref{nl_Ohms_law}, \ref{diverge_j}, \ref{diverge_j0_1} and \ref{diverge_j_2} yields a compact eigenmode equation, termed the `RSAE-ZF-MHD' model:
\begin{equation}\label{AE_ZF_eigen_eq}
{\bm\nabla}_\perp\cdot(\epsilon_A{\bm\nabla}_\perp\delta\phi_0)+\Lambda\delta\phi_0+{\bm\nabla}\cdot \left[\left({\bm\nabla}\alpha_{\phi 2}\right)\delta\phi_0\right]=0,
\end{equation}
where
\begin{equation}\label{AE_ZF_eigen_eq_components}
\begin{aligned}
&\epsilon_A=\epsilon_{A0}+\alpha_{\phi 1}+\alpha_A, \\
&\epsilon_{A0}=\left[\left(1-\omega_{\ast pi}/\omega_0\right)(\omega_0^2-\omega_{\rm gam}^2)\right]-k_{\parallel 0}^2\upsilon_{\rm A}^2, \\
&\alpha_{\phi 1}=\omega_0\frac{c}{B_0}k_{\theta 0}\left[\left(\frac{k_{\parallel 0}\upsilon_{\rm A}}{\omega_0}\right)^2+1-\frac{\omega_{\ast pi}}{\omega_0}\right]\left(\frac{\partial\delta\phi_z}{\partial r}\right), \\
&\alpha_A=-2\omega_0\frac{c}{B_0}k_{\theta 0} \left(\frac{k_{\parallel 0}\upsilon_{\rm A}}{\omega_0}\right) \left(\frac{\upsilon_{\rm A}}{c}\right)\left(\frac{\partial\delta A_{\parallel z}}{\partial r}\right), \\
&\alpha_{\phi 2}=\omega_0\frac{c}{B_0}k_{\theta 0}\left[\left(\frac{k_{\parallel 0}\upsilon_{\rm A}}{\omega_0}\right)^2-1+\frac{\omega_{\ast pi}}{\omega_0}\right]\left(\frac{\partial\delta\phi_z}{\partial r}\right).
\end{aligned}
\end{equation}
Here, $\Lambda$ incorporates effects from favorable average curvature, toroidal couplings, as well as wave-particle interactions due to thermal and EPs \cite{Zonca2002,Yu2009}. The potentials $\delta\phi_z$ and $\delta A_{\parallel z}$ yield, respectively, the zonal flow and zonal current, which are beat-driven by the RSAEs as detailed in Ref. \cite{Chen2025}.

{\it Variational Analysis}.$-$We perform a variational analysis of the `RSAE-ZF-MHD' eigenmode equation, Eq. \ref{AE_ZF_eigen_eq}, for a RSAE localized near $r=r_{\rm m}$, using the normalized radial coordinate $\zeta=k_{\theta 0}(r-r_{\rm m})$ with $k_{\theta 0}=m_0/r_{\rm m}$. Introducing $\psi=\epsilon_A^{1/2}\delta\phi_0$, Eq. \ref{AE_ZF_eigen_eq} leads to the variational form:
\begin{equation}\label{AE_ZFs_variational_form}
\begin{aligned}
{\cal L}\left[\psi\right]=\frac{1}{2}\int_{-\infty}^{\infty}d\zeta\bigg\{&-\left|\frac{d\psi}{d\zeta}\right|^2
+\left|\psi\right|^2\bigg[-1+\frac{1}{4}\left(\frac{{\hat\epsilon'}_A}{{\hat\epsilon}_A}\right)^2 \\
&-\frac{1}{2}\left(\frac{{\hat\epsilon''}_A}{{\hat\epsilon}_A}\right)
+\frac{{\hat\Lambda}}{{\hat\epsilon}_A}+\frac{1}{2}\frac{\alpha''_{\phi 2}}{{\hat\epsilon}_A}\bigg]\bigg\},
\end{aligned}
\end{equation}
where ${\hat\epsilon}_A={\hat\epsilon}_{A0}+{\hat\alpha}_{\phi 1}+{\hat\alpha}_A$, ${\hat\epsilon}_{A0}=(1-\Omega_{\ast pi}/\Omega_0)(\Omega_0^2-\Omega_{\rm gam}^2)-K_{\parallel 0}^2$, $\Omega_0=\omega_0/\omega_{A}$ with $\omega_{A}=\upsilon_{A}/(q_{\rm m}R)$ and same normalization adopted for other frequency variables, $K_{\parallel 0}=(n_0q-m_0)$, ${\hat\Lambda}=\Lambda/(k_{\theta 0}\omega_A)^2$, $g'\equiv dg/d\zeta$, and all other hatted quantities are normalized by $\omega_{A}$.

For $n_0q_{\rm m}\gg1$, the RSAE is strongly localized around $\zeta=0$, allowing the approximation $K_{\parallel 0}\simeq K_{\rm m}+\frac{1}{2}\frac{S^2}{n_0q_{\rm m}}\zeta^2$, with $K_{\rm m}=(n_0q_{\rm m}-m_0)<0$ and $S^2\equiv r_{\rm m}^2(d^2q/dr^2)/q_{\rm m}$. This yields $\epsilon_{A0}\simeq \frac{|K_{\rm m}|S^2}{n_0q_{\rm m}}(\zeta^2+\zeta_0^2)$ with $\zeta_0^2=[(1-\Omega_{\ast pi}/\Omega_0)(\Omega_0^2-\Omega_{\rm gam}^2)-K_{\rm m}^2](n_0q_{\rm m})/(S^2|K_{\rm m}|)$. Using the trial function $\delta\psi_t=\exp(-|\zeta|)$, Eq. \ref{AE_ZFs_variational_form} then readily yields the linear RSAE variational disperison relation without ZFs:
\begin{equation}\label{linear_rsae_variational_dr}
\zeta_0=(\pi/2)\big[{\hat\Lambda}n_0q_{\rm m}/\big(|K_{\rm m}|S^2\big)-1/2\big],
\end{equation}
in agreement with previous analytic results \cite{Berk1993a,Zonca2002}.

Continuum damping can be incorporated by evaluating the Alfv\'en resonant absorption near $\zeta=\pm \zeta_2$, where $\epsilon_{A0}=0$. Near the resonance, $\epsilon_{A0}\simeq -[(1-\Omega_{\ast pi}/\Omega_0)(\Omega_0^2-\Omega^2_{\rm gam})]^{1/2}(S^2/n_0q_{\rm m})(\zeta^2-\zeta_2^2)$ with $\zeta_2^2=(2n_0q_{\rm m}/S^2)\{[(1-\Omega_{\ast pi}/\Omega_0)(\Omega_0^2-\Omega_{\rm gam}^2)]^{1/2}+|K_{\rm m}|\}$. The modified dispersion relation including continuum resonant-absorption damping becomes:
\begin{equation}\label{linear_rsae_variational_dr_with_reson}
\zeta_0(1+i\delta_c)=(\pi/2)\big[{\hat\Lambda n_0q_{\rm m}}/\big(|K_{\rm m}|S^2\big)-1/2\big],
\end{equation}
where $\delta_c=(\pi/2)\exp(-2\zeta_2)$ is the continuum damping rate, and ${\mathbb R}e(\zeta_0), {\mathbb R}e(\zeta_2)>0$. This result, which recovers established analytical theory \cite{Zonca2002}, shows that $\delta_c$ is typically negligible due to the exponential factor \cite{Deng2012}.

Extending the same approach to include ZFs, we introduce $\alpha_z={\hat\alpha}_{\phi 1}+{\hat\alpha}_A$ and expand around $\zeta=0$, obtaining ${\hat\epsilon}_A={\varDelta}_z\left(\zeta^2+\zeta_{0z}^2\right)$ with:
\begin{equation}\label{Delta_z_with_zfs}
\begin{aligned}
 &\varDelta_z=\frac{1}{(n_0q_{\rm m})^2}\left[\left|K_{\rm m}\right|n_0\left(r^2\frac{d^2q}{dr^2}\right)_{\rm m} +\frac{1}{2}\left(r^2\frac{d^2\alpha_z}{dr^2}\right)_{\rm m}\right], \\
 &\zeta_{0z}^2=\left[\left(1-\Omega_{\ast pi}/\Omega_0\right)\left(\Omega_0^2-\Omega_{\rm gam}^2\right)-K_{\rm m}^2+\alpha_{z{\rm m}}\right]/\varDelta_z,
\end{aligned}
\end{equation}
where the subscript `${\rm m}$' denotes evaluation at $r=r_{\rm m}$. Using the same trial function $\delta\psi_t=\exp(-|\zeta|)$ yields the RSAE-ZFs variational dispersion relation:
\begin{equation}\label{linear_rsae_variational_dr_with_zfs}
\zeta_{0z}\left(1+i\delta_c\right)=\left(\pi/2\right)\left({\hat\Lambda}_z/\varDelta_z-1/2\right),
\end{equation}
with ${\hat\Lambda}_z={\hat\Lambda}+(d^2{\hat\alpha}_{\phi 2}/dr^2)_{\rm m}/2$. This constitutes the general RSAE dispersion relation including ZF effects.

The self-consistent ZFs, which are beat-driven by the RSAE \cite{Chen2025}, enable explicit evaluation of the nonlinear terms, that is, $\alpha_{z{\rm m}}\simeq -2(\frac{c}{B_0\omega_{A}})^2(\frac{n_0q}{r})^2_{\rm m}\{(\frac{K_{\parallel 0}}{\Omega_0})^2-\frac{1}{2}\frac{\Omega_{\ast pi}}{\Omega_0}[(\frac{K_{\parallel 0}}{\Omega_0})^2+1-\frac{\Omega_{\ast pi}}{\Omega_0}]\}_{\rm m}(\frac{\partial^2|\Phi_0|^2}{\partial\zeta^2})_{\rm m}$ and
$\frac{d^2{\hat\alpha}_{\phi 2}}{dr^2}|_{\rm m}\simeq (\frac{c}{B_0\omega_{A}})^2 (\frac{n_0q}{r})_{\rm m}^4 (\frac{\Omega_{\ast pi}}{\Omega_0})[(\frac{K_{\parallel 0}}{\Omega_0})^2 -1 +\frac{\Omega_{\ast pi}}{\Omega_0}]_{\rm m}(\frac{\partial^2|\Phi_0|^4}{\partial\zeta^4})_{\rm m}$.
Under the typical ordering $|\Omega_0|\approx|K_{\rm m}|>|\Omega_{\ast pi}|$ with $|\Phi_0|^2$ peaking at $\zeta=0$, we find $\alpha_{z{\rm m}}>0$. Equations \ref{Delta_z_with_zfs} and \ref{linear_rsae_variational_dr_with_zfs} thus predict a net downward frequency shift of the RSAE ($\Omega_{0r}$) with increasing ZFs. By direct inspection of the nonlinear terms, it is possible to verify that zonal current ($\delta A_{\parallel z}$) dominates over the subdominant zonal flow ($\delta\phi_z$), which by itself would induce an opposite upward frequency shift.

This analysis predicts that ZF-induced downward frequency chirping shifts the RSAE frequency toward the accumulation point of the continuum spectrum, increasing the effective radial wavenumber $k_r$ and enhancing mode conversion to KAWs. To self-consistently capture this physics, we extend the our theoretical framework to include the KAW dynamics and electron Landau damping, the latter effect being missing in the numerical description adopted in current GTC simulations.

{\it RSAE-ZF-KAW Model}.$-$For the coupled RSAE-KAW system where electron Landau resonance becomes relevant ($|\omega_0|\sim|k_\parallel\upsilon_{te}|$), the non-adiabatic electron response is give by
\begin{equation}\label{delta_g_e_RSAE-KAW}
\begin{aligned}
\delta g_e\simeq -\frac{e}{T_e}\left(1-\frac{\omega_{\ast e}}{\omega_0}\right)F_{Me}\left(\delta\psi_0-\frac{\omega_0}{k_\parallel\upsilon_{te}-\omega_0}\delta\phi_\parallel\right),
\end{aligned}
\end{equation}
where $\delta\phi_\parallel\equiv \delta\phi_0-\delta\psi_0$. The ion response remains unchanged from the MHD description and is given by $\delta g_{i0}^{(1)}$ in Eq. \ref{delta_g_i0_1_2}. Substituting these responses into the quasi-neutrality condition yields $\delta\psi_0={\hat\sigma}_k\delta\phi_0$, where ${\hat\sigma}_k=1+\tau(1-\Gamma_0)(1-\omega_{\ast i}/\omega_0)\big/[(1-\omega_{\ast e}/\omega_0)(1+\zeta_eZ(\zeta_e))]$ incorporates FLR effects via $\Gamma_0=I_0(b_i)\exp(-b_i)$ and electron kinetics via the plasma dispersion function $Z(\zeta_e)$, with $\zeta_e=\omega_0/(|k_\parallel|\upsilon_{te})$. Here, $\tau=T_e/T_i$, $I_0$ is the zeroth-order modified Bessel function, and $b_i=k_\perp^2\rho_i^2$.

Assuming $|k_\perp\rho_i|^2 < 1$, the nonlinear gyrokinetic vorticity equation \cite{Chen2016} then yields, after some straightforward manipulations, the fourth-order eigenmode equation for the `RSAE-ZF-KAW' model:
\begin{equation}\label{RSAE_KAW_e_Landau_model}
\begin{aligned}
\rho_i^2\nabla_\perp^2\big(1-\omega_{\ast pi}/\omega_0\big)&\Big\{\frac{3}{4}\left(\omega_0^2-\omega_{\rm gam}^2\right)\\
+\tau k_\parallel^2\upsilon_{\rm A}^2 \big/\big[(1&-\omega_{\ast e}/\omega_0)(1+\zeta_eZ(\zeta_e))\big]\Big\}\nabla_\perp^2\delta\phi_0 \\
+{\bm\nabla}_\perp\cdot(\epsilon_{A}{\bm\nabla}_\perp\delta\phi_0)&+\Lambda\delta\phi_0 +{\bm\nabla}\cdot \left[\left({\bm\nabla}\alpha_{\phi 2}\right)\delta\phi_0\right]=0.
\end{aligned}
\end{equation}
This extended model self-consistently describes the coupling between RSAEs, ZFs, and KAWs, incorporating both FLR effects and electron Landau damping. The FLR terms not only regularize the singularity at $\epsilon_{A}=0$ present in the MHD description, Eq. \ref{AE_ZF_eigen_eq}, but also capture the essential physics of mode conversion from RSAEs to KAWs. Thereby, the model provides a foundation for analyzing the enhanced radiative damping responsible for the nonlinear suppression and eventual saturation of the RSAE. Derivation of Eq. \ref{RSAE_KAW_e_Landau_model} is particularly motivated by the limitations of the model for electron response implemented in current GTC simulations. While accurately describing short-time saturation via convective damping with a fluid-kinetic hybrid electron model \cite{Lin2001}, approximations implied by this approach cannot capture the long-time global dynamics of KAW physics. The present analytic theory, thus, complements the GTC simulations by enabling the study of this crucial physics.

{\it Numerical Validations}.$-$We present numerical solutions of the `RSAE-ZF-MHD' and `RSAE-ZF-KAW' models to quantitatively validate theoretical predictions and distinguish the roles of zonal flow and zonal current, with direct comparison to those of gyrokinetic simulations.

\begin{figure}[htbp]
\centering
\includegraphics[width=0.36\textwidth]{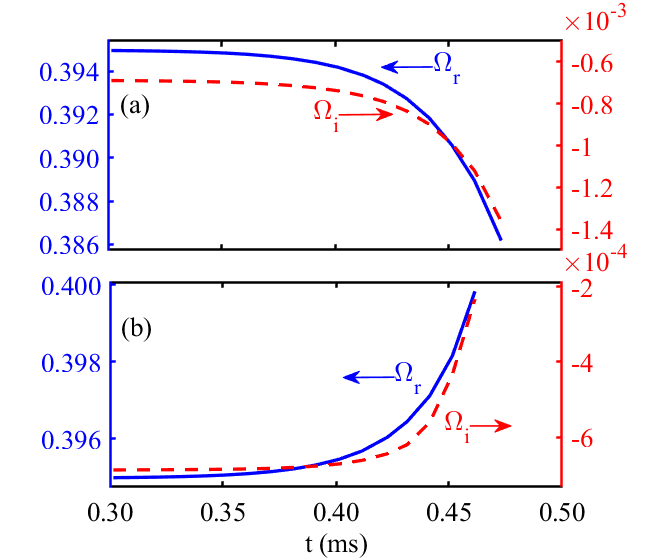}
\caption{(a) Time evolution of normalized frequency $\Omega_r=\omega_{0r}/\omega_A$ and damping rate $\Omega_i=\omega_{0i}/\omega_A$ from the `RSAE-ZF-MHD' model with zonal flow and zonal current. (b) Corresponding evolution with only zonal flow included.}
\label{RSAE_MHD_full_zfs}
\end{figure}

Eigenmode solutions confirm the theoretical prediction of zonal current dominance. When the both zonal components are included, the `RSAE-ZF-MHD' model exhibits clear downward chirping (FIG. \ref{RSAE_MHD_full_zfs}(a), blue). Artificially removing zonal current reverses this trend, yielding a smaller upward frequency chirping (FIG. \ref{RSAE_MHD_full_zfs}(b), blue); thus, demonstrating that zonal current dominates the nonlinear frequency shift. In both cases, continuum damping remains negligible, of the order of $10^{-4}$ to $10^{-3}$ (FIGs. \ref{RSAE_MHD_full_zfs}(a) and (b), red), in excellent agreement with theoretical expectations.

\begin{figure}[htbp]
\centering
\includegraphics[width=0.35\textwidth]{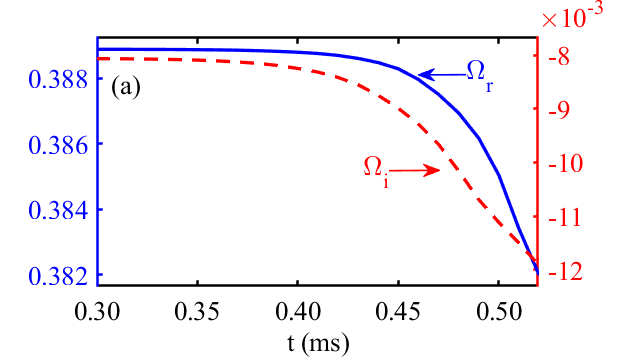}
\includegraphics[width=0.35\textwidth]{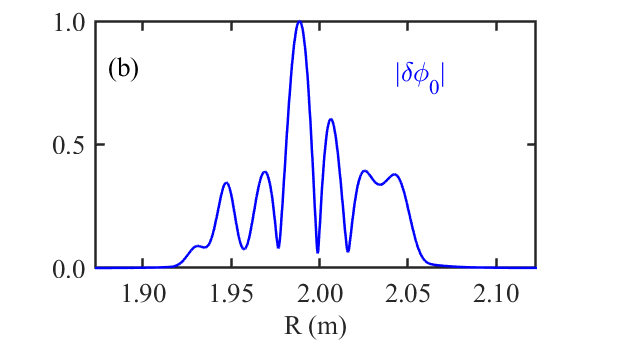}
\caption{(a) Time evolution of $\Omega_r$ and $\Omega_i$ from the `RSAE-ZF-KAW' model with zonal flow and zonal current. (b) Radial profile of the perturbed electrostatic potential $|\delta\phi_0|$ at the nonlinear saturation ($t=0.53$ ms).}
\label{RSAE_KAW_ZFs}
\end{figure}

Radiative damping via KAWs is identified as the primary mechanism by which ZFs yield RSAE saturation. This conclusion is supported by three key findings from the `RSAE-ZF-KAW' model: (1) it reproduces the downward frequency chirping and yields a radiative damping rate nearly two orders of magnitude stronger than MHD continuum damping (FIG. \ref{RSAE_KAW_ZFs}(a)); (2) it predicts the development of fine-scale radial structures in the perturbed electrostatic potential (FIG. \ref{RSAE_KAW_ZFs}(b)), confirming the increase in radial wavenumber $k_r$ predicted by the eigenmode theory and in agreement with GTC observations (FIG. \ref{gtc_sim_fig2}(a)); (3) most compellingly, it captures a rapid, order-of-magnitude increase in $E_\parallel/E_{\parallel,{\rm es}}$ (FIG. \ref{RSAE_KAW_deltaE_parallel}, red squares), in excellent agreement, again, with GTC simulations both qualitatively and quantitatively (FIG. \ref{RSAE_KAW_deltaE_parallel}, black). This enhancement is a clear signature of KAW excitation, which the `RSAE-ZF-MHD' model (FIG. \ref{RSAE_KAW_deltaE_parallel}, blue circles) fails to describe as the RSAE merges into the Alfv\'en continuous spectrum. In this case, the ZF contribution to $E_\parallel$ remains over an order of magnitude smaller throughout, confirming KAW generation as the primary mechanism responsible for the strong enhancement of $E_\parallel$.

{\it Conclusions and Discussions}.$-$This study establishes a comprehensive physical picture of RSAE suppression and saturation via the thermal plasma nonlinearity route. Combining gyrokinetic simulations and analytical theory, we demonstrate that zonal current$-$not zonal flow$-$dominates the nonlinear frequency chirping, and that convective (radiative) damping via mode-converted KAWs$-$not continuum damping$-$governs the energy dissipation. The developed `RSAE-ZF-KAW' model qualitatively and quantitatively captures this physics and provides a crucial extension of the current GTC simulation framework by incorporating the full electron kinetics necessary for describing long-time global KAW dynamics.

Our study suggests that one may expect the processes described to be generally applicable to Alfv\'en eigenmode dynamics. They not only illuminate the key ambiguity regarding the role of zonal electromagnetic fields in the suppression and saturation of Alfv\'enic fluctuations, but also unveil a novel energy transfer channel$-$from unstable EP-driven-eigenmodes to damped KAWs, mediated by zonal current. This paradigm, centered on zonal-current-induced frequency chirping and subsequent enhanced radiative damping, provides a fundamental framework for understanding Alfv\'en eigenmode nonlinear evolution in toroidal plasmas. Furthermore, the identified energy channeling process opens up new avenues for future research into alpha-particle transport in burning plasmas.


This work has been supported by the National key R\&D Program of China (Grant Nos. 2022YFE03040002 and 2024YFE03170000), the National Science Foundation of China (Grant Nos. 12175053, 12405265 and 12261131622), the Italian Ministry of Foreign Affairs (Grant No. CN23GR02), the Natural
Science Foundation of Sichuan (Grant No. 2025ZNSFSC0064), the Strategic Priority Research Program of the Chinese Academy of Sciences (Grant No. XDB0500302), and the start-up funding of Institute of Physics, Chinese Academy of Sciences (Grant No. E3CB031R21). Part of this work has been carried out within the framework of the EUROfusion Consortium, funded by the European Union via the Euratom Research and Training Programme (Grant Agreement No 101052200$-$EUROfusion).

\section*{References}
$^\ast$ $\text{Corresponding author: liuchen@zju.edu.cn}$


\end{document}